\documentclass[12pt,preprint]{aastex}

\shorttitle{No New Class of GRBs}
\shortauthors{Schaefer}

\begin{document}

\title{GRB060614 is at High Redshift, So No New Class of Gamma-Ray Bursts Is Required}

\author{Bradley E. Schaefer}
\author{Limin Xiao}
\affil{Physics and Astronomy, Louisiana State University,
    Baton Rouge, LA, 70803}

\begin{abstract}

Long duration Gamma-Ray Bursts (GRBs) have been strongly connected with core collapse supernovae, so it was surprising when the recent GRB060614 (with a reported redshift of 0.125) was found to have no visible supernova to deep limits.  Three separate groups have reached the same conclusion that this event forces the existence of a new previously-unsuspected class of GRBs.  The problem with this conclusion is that the redshift is not secure, as the measured galaxy emission lines simply give the redshift of the brightest galaxy near the line-of-sight to the GRB afterglow.  Fortunately, eight different luminosity indicators are known which can give the luminosity (and hence redshift) of the GRB itself.  In combination, these luminosity indicators have already proven to give redshifts with average one-sigma errors of 26\% as based on 69 GRBs with spectroscopic redshifts.  For GRB060614, the luminosity indicators all uniformly give a high luminosity for the burst which implies a high redshift.  This can be seen by the spiky many-peaked light curve, the high $E_{peak}$ value, the near-zero spectral lag, and the timing of the jet break.  Quantitatively, we derive a redshift of $1.97_{-0.53}^{+0.84}$ from the eight luminosity indicators.  Additional information from the detection of the afterglow in all {\it Swift UVOT} bands further constrains the redshift to be $<1.71$, so the final range of allowed redshift is $1.44<z<1.71$, which readily explains why no supernova was visible.  Thus, GRB060614 is at high redshift and is certainly not evidence for a new class of GRBs.

\end{abstract}



\section{Introduction}

	Long-duration Gamma-Ray Bursts (GRBs) are now known to be associated with Type Ib/c supernovae.  The primary evidence for this is four identifications of spectroscopically-confirmed supernovae with core collapse at the same time and place as the GRBs (Galama et al. 1998; Hjorth et al. 2003; Stanek, K. Z. et al. 2003; Malesani, D. et al. 2004; Campana et al. 2006).  Along with secondary evidence (e.g., Schaefer et al. 2003) and strong theoretical support (e.g., MacFadyen, Woosley, \& Heger 2001; Woosley \& Bloom 2006), the GRB community is now confident that GRBs arise promptly from the core collapse of supernovae.  A prediction of this new paradigm is that all nearby GRBs should have visible supernovae.
	
	After GRB060614 exploded, a redshift of 0.125 was determined (Price et al. 2006) and this event became a test of the paradigm prediction.  Three groups (Della Valle et al. 2006; Gal-Yam et al. 2006; Fynbo et al. 2006) tested this prediction and discovered that no supernova was visible to very strict limits.  This contradiction to the current paradigm forced all three groups to conclude that there must be a separate eruption mechanism different from the standard model.  Indeed, the titles of their three papers declare that "GRB060614 requires a novel explosion process", that GRB060614 is "not due to a hypernova", and that it represents "a new type of massive stellar death".

	The weak point in the case for a new class of GRBs is the distance to GRB060614.  The reported wavelengths of the observed galaxy emission lines only gives the redshift of the brightest galaxy near the line of sight to the transient.  What is needed is a redshift for the GRB.  And exactly this can be provided with all needed accuracy by any of eight luminosity indicators.  The purpose of this paper is to determine the redshift to GRB060614 and test whether the lack of a visible supernova is simply due to the burst really being at high redshift.
	
\section{Luminosity and Redshift of GRB060614}

	Since 1999, a total of eight different luminosity indicators for GRBs have been discovered (Norris, Marani, \& Bonnell 2000; Fenimore \& Ramirez-Ruiz 2000; Schaefer 2003; Amati et al. 2002; Ghirlanda, Ghisellini, \& Lazzati 2004; Schaefer 2002; Firmani et al. 2006).  These indicators are measurable photometric and spectroscopic properties of the burst which are correlated as power laws with the burst luminosity and energy.  With this, we can measure the indicators for any burst, deduce the luminosity, combine the luminosity with the observed peak flux and the inverse-square law of light to get the luminosity distance, and then use some fiducial cosmology to find the redshift of the burst.  There are reasonable understandings for the physical bases of these luminosity relations (Schaefer 2004; M\'{e}sz\'{a}ros et al. 2002; Kobayashi, Ryde, \& MacFadyen 2002; Schaefer 2003; Rees \& M\'{e}sz\'{a}ros 2005; Schaefer 2002; Thompson, M\'{e}sz\'{a}ros, \& Rees 2006).  These relations give luminosities of varying accuracy, and when combined give accuracies that are surprisingly good.  From our sample of 69 GRBs with spectroscopic redshifts, our deduced redshifts have a one-sigma scatter of 26\% about the known values.  Figure 1 presents our 69 redshifts (with one-sigma error bars) plotted versus  the spectroscopic redshifts.  The reduced chi-square of the comparison between our derived and spectroscopic redshifts is close to unity, demonstrating that our calculated uncertainties are reasonable.  With this, we have a tool for measuring the redshift to GRB060614.
	
	GRB060614 is a particularly good case for measuring the needed luminosity indicators for all eight known luminosity relations.  This is because the {\it Swift} light curve has good signal (see Figure 2), the {\it Konus-Wind} experiment detected the burst to high energy so as to pin down the photon energy of the spectral peak ($E_{peak}$), and ground-based photometry revealed a jet break in the afterglow light curve.  Thus, we can derive eight measures of the GRB distance independent of the redshift of the brightest galaxy along the line of sight.  Table 1 itemizes the eight luminosity relations and their derived luminosities.  Six of the luminosity relations are as calibrated by ourselves with 69 GRBs with spectroscopic redshifts, while the fourth relation is taken from Amati (2006) and the last relation is taken from Firmani et al. (2006).
	
	From the {\it Swift} data archives, we extracted the light curves for GRB060614 for energy ranges 25-350 (see Figure 2), 25-50, and 100-350 keV with 0.064 second time resolution.  The spectral lag, $\tau _{lag}$, is the offset of the peak in the cross correlation function between the 25-50 and 100-350 keV light curves.  For GRB060614, this cross correlation has a well defined peak so that $\tau_{lag}=0.02 \pm 0.02$ seconds.  The variability, $V$, is the normalized RMS scatter of the observed 25-350 keV light curve about a smoothed version of the same light curve.  For the particular optimized formulation used by us as calibrated for 69 GRBs with known redshifts, we find $V=0.0050 \pm 0.0002$.  The minimum rise time, $\tau_{RT,min}$, is taken as the shortest time in the 25-350 keV light curve that the brightness rises by 50\% of the flux for the associated peak.  In the case of GRB060614, the highest bin is immediately preceded by a bin with a little less than half the peak flux, which yields $\tau_{RT,min}=0.06 \pm 0.01$ seconds.  The number of peaks, $N_{peak}$, is taken as the number of local maxima in the 25-350 keV light curve that are higher than 25\% of the peak flux and which are separated from other peaks by a minimum that is at least 25\% of the peak flux below the surrounding peaks.  GRB060614 has five sharp and high peaks in its first five seconds followed by four broad and relatively low peaks in the next minute, for $N_{peak}=9$.  The duration $T_{0.45}$ is defined by Reichart et al. (2001) as the cumulative time over which the brightest bins include 45\% of the burst fluence.  With the 25-350 keV light curve, we find $T_{0.45}=18 \pm 1$ seconds.  The photon energy of the peak spectral flux, $E_{peak}$, is given by Golenetskii et al. (2006) to be $302_{-85}^{+214}$ keV as based on the {\it KONUS-Wind} spectrum from 20-2000 keV.  Della Valle et al. (2006) observed a significant break in the light curve at $1.38 \pm 0.04$ days that they claim is the jet break.  With the usual assumptions of a uniform circumstellar medium density of 3 cm$^{-3}$ and an efficiency of 0.2 (cf. Nava et al. 2006), we derive a half-opening angle for the jet of $3.16\degr \pm 0.05\degr$ and a beaming factor, $F_{beaming}$, of $0.0015 \pm 0.0001$ for low redshifts (and somewhat smaller for higher redshifts).
	
	We quickly see that the luminosity indicators all uniformly and strongly give a high luminosity for GRB060614.  The $\tau _{lag}$ value is near zero giving a very high luminosity, the $V$ value is large (i.e., the light curve is spiky; see Figure 2) indicating a high luminosity, the $E_{peak}$ value is relatively high again pointing to a high luminosity (Pelangeon \& Atteia 2006), the early jet break time yields a high burst energy, the $\tau _{RT,min}$ is very short also suggesting a high luminosity, and the $N_{peak}=9$ result (see Figure 2) puts a high lower-limit on the luminosity.
	
	To be quantitative, we have taken the measured luminosity indicators and derived the luminosities as given in Table 1.  The derived values are $L$ (the isotropic peak luminosity in units of erg s$^{-1}$), $E_{\gamma ,iso}$ (the isotropic burst energy in $\gamma$ radiation in units of ergs), and $E_{\gamma}$ (the burst energy in $\gamma$ radiation corrected for the effects of beaming in units of ergs).  The derived luminosities depend on the redshift of the burst because the time scales are corrected for dilation and the $E_{peak}$ is corrected for redshift back to the GRB rest frame.  Therefore, we present the derived luminosities for two redshifts; z=0.125 to represent the nearby case and z=1.97 to represent the optimal solution.
	
	Barthelmy et al. (2006) give the 1-second peak flux as $11.6 \pm 0.4$ photon cm$^{-2}$ s$^{-1}$ in the 15-150 keV energy band with one-sigma error bars.  Golenetskii et al. (2006) give the fluence to be $4.09_{-0.34}^{+0.18} \times 10^{-5}$ erg cm$^{-2}$ in the 20-2000 keV energy band.  With a smoothly broken power law (Band et al. 1993), the spectral parameters from Golenetskii et al. (2006), and the average high-energy power law index of $-2.2 \pm 0.4$, we can derive the bolometric peak flux and fluence.  Here, bolometric is taken to be from 1-10,000 keV because this contains essentially all the flux.  We find the bolometric peak flux is $3.0 \pm 0.3 \times 10^{-6}$ photon cm$^{-2}$ s$^{-1}$ and the bolometric fluence is $1.5 \pm 0.2 \times 10^{-4}$ erg cm$^{-2}$.  With the inverse-square law for light plus an assumed distance, we can convert the bolometric peak flux into $L$ and the bolometric fluence into $E_{\gamma ,iso}$.  With the beaming factor ($F_{beaming}$), we can covert $E_{\gamma ,iso}$ into $E_{\gamma}$.  For converting from luminosity distance to redshift, we will be adopting the concordance cosmology with $\Omega _M=0.27$ in a flat universe with an unchanging cosmological constant of $w=-1$.
	
	{\it If} the GRB is at a redshift of 0.125, the observed bolometric peak flux corresponds to $\log L = 50.04$.  This is greatly fainter than all the derived $\log L$ values from the luminosity indicators (see the middle column of Table 1).  At a redshift of 1.97, the observed $\log L$ is $52.94$, which is right in the middle of the observed values (see the last column of Table 1).  This quantitative result is just repeating again that GRB060614 is at high redshift (and hence simply explains the lack of any visible supernova).
	
	To derive the optimal value for the redshift of GRB060614, we use a chi-square comparison between the observed and derived luminosities as a function of redshift.  The minimum chi-square is 4.15 at a redshift of z=1.97.  (The number of degrees of freedom should be 6, with 7 relations [because the $N_{peak}$ relation only returning a limit] minus 1 for the single fit parameter of $z$.  The eight relations are not entirely independent, for example with the $E_{peak}-E_{\gamma, iso}$ relation just being a poor version of the $E_{peak}-E_{\gamma}$ relation, so the effective number of degrees of freedom will be somewhat smaller than 6.  Thus, the reduced chi-square is of order unity, and this tells us that all the luminosity indicators are consistent.)  The one-sigma confidence region is where the chi-square value is within 1.0 of this minimum, which is $1.44<z<2.81$.  The chi-square for z=0.125 is 315.  
	
	We have additional information on the redshift from the positive detection of the GRB060614 afterglow in all {\it Swift UVOT} bands (Holland et al. 2006).  The UVW2 filter has maximum sensitivity around 0.2 $\mu$ yet with response out to past 0.3 $\mu$, allowing detections out to redshifts approaching 2 at least in principle.  In practice, GRB050802 is certainly at a redshift of 1.71 (Fynbo et al. 2005) despite being detected in all {\it Swift UVOT} bands (McGowan et al. 2005).  This implies that the redshift range from the luminosity indicators has its high end disallowed by the {\it UVOT} data, with the upper limit being roughly $z=1.71$.  The lack of detected metal absorption lines in the GRB060614 afterglow (Fugazza et al. 2006) can be readily explained by the even chance that the GRB is on the nearside of its host galaxy or by precedent (Chen et al. 2006).  (Metal absorption lines are frequent in GRBs with {\it known} redshifts $\sim 1.5$, but this is no statement about the frequency of $z \sim 1.5$ bursts with no such lines as their redshifts would not be determined.)  In all, the additional optical and ultraviolet data only places a constraint on the redshift to be less than something like 1.71, and this is fully consistent with our independent redshift measurement.
	
	Thus, the redshift for GRB060614 is $1.44<z<1.71$ and is certainly not 0.125.

\section{Coincidence Probability}
		
	With GRB060614 being at high redshift, we have to accept that the alleged host galaxy falls near the line-of-sight to the afterglow by chance alone.  Does this condition require a coincidence that is far-fetched?  That is, what is the expected number of false alarms caused by nearby galaxies being close to the line of sight towards any GRB optical transient?  
	
	For this probability calculation, the real question is what would be accepted as a positional coincidence and published as evidence of a causal connection?  In the three papers (Fynbo et al. 2006; Della Valle et al. 2006; Gal-Yam et al. 2006), they were willing to accept angular separations as large as 4.3 arc-seconds (as for GRB060505, see below) and galaxy luminosities brighter than the alleged host galaxy ($M_B=-15.6$; Fynbo et al. 2006).  Further, for false alarms, the intervening galaxy must be closer than roughly $z=0.2$ for strong limits on the supernova to be placed.  A cone with a half-opening angle of 4.3$\arcsec$ and a height corresponding to z=0.2 has a volume of 0.4 Mpc$^3$.  For $M_B^*=-19.5$ (Trimble 2000), a galaxy could provide a false alarm if its luminosity is greater than $0.03 \times L^*$.  For the Schechter luminosity function normalized to the average galaxy density in the local Universe with $h=0.7$ (Trimble 2000), the integral over the luminosity function for galaxies brighter than $0.03 \times L^*$ is 0.02 galaxies per cubic Megaparsec.  So the expectation value for the number of false-alarm nearby galaxies is 0.008 for a single burst.  In other words, with 125 random positions on the sky, we expect that roughly one nearby galaxy would give rise to a false alarm about a missing GRB supernova.
	
	To date, a total of 143 GRB optical afterglows have been discovered (Greiner 2006), any one of which could have had a $z<0.2$ galaxy near the line-of-sight.  With this, the expected number of false alarms is somewhat greater than one.  We would not be surprised at two false alarms.  This quantitative result has substantial uncertainty due to the normal variations in threshold for what different observers would accept as a positional coincidence.  Nevertheless, it is clear that the number of expected false alarms is similar to the the number of claimed nearby host galaxies with no supernovae.

\section{GRB060505}

	GRB060505 has been identified by Fynbo et al. (2006) as a second case of a low redshift event with severe limits on any present supernova event.  
	
	GRB060505 was discovered by {\it Swift} only by analysis on the ground many hours after the event, because the {\it BAT} onboard image brightness was below the required significance threshold. (Palmer et al. 2006).  This faint brightness is in part caused by the far-off-axis position of the burst, with the partial coding being only 11\% (Hullinger et al. 2006).  In addition, the four-second long burst occurred during a time interval with a very fast rising background due to the spacecraft approaching the South Atlantic Anomaly.  The {\it Swift XRT} instrument then detected a faint and fading x-ray source in the $\gamma$-ray error circle (Conciatore et al. 2006).  Thus, the circumstances conspired to make this a faint and poorly observed burst.
	
	Two groups (Ofek et al. 2006; Fynbo et al. 2006) identified a fading afterglow, both apparently with a single early image.  The afterglow position is 4.3$\arcsec$ from the center of a galaxy with a confidently measured redshift of 0.089 (Ofek et al. 2006).  This position is far outside virtually all the flux from the galaxy, although Fynbo et al. (2006) reports that an apparent spiral arm does extend out to the afterglow position.  Subsequent monitoring shows the lack of any supernova light with limits 2-4 magnitudes fainter than expected for previously known GRB supernovae placed at a redshift of 0.089 (Fynbo et al. 2006).
	
	Unfortunately, the poor observing circumstances do not allow for any of the luminosity indicators to be usefully measured.  The high noise level and far-off-axis conditions combined with a faint burst to produce a light curve with the noise dominating over the signal.  Thus, the 100-350 keV band has no significant flux for us to calculate $\tau _{lag}$, the uncertainty from the subtraction of the Poisson contribution to $V$ is much larger than any expected signal, the light curve must be heavily binned so that the rising signal dominates over the Poisson noise with the result that there is too poor a time resolution for measuring $\tau _{RT,min}$ over the relevant range, and the one apparent peak in the light curve is ambiguous as to the burst's luminosity.  The faint recorded flux plus the lack of detections by other satellites means that no $E_{peak}$ is known.  The existence of only two positive detections of the afterglow means that no jet break time is known.  In all, GRB060505 is one of the worst possible cases for extracting luminosity indicators, so we cannot get any useful estimate of its redshift.
	
	Again, the weak point in the case for GRB060505 requiring a new class of GRBs is the redshift of the burst.  The only evidence for the burst being at z=0.089 is that there is a z=0.089 galaxy near the line of sight to the afterglow.  But, as the previous section shows, the false alarm rate leads us to {\it expect} one-to-two spurious missing GRB supernovae to date.  So we are left with trying to choose between the hypothesis that an entirely new class of GRBs must be invented or the hypothesis that the z=0.089 galaxy is a chance foreground interloper exactly as expected.  Occam's Razor strongly comes down on the side of choosing the known and expected versus postulating a new phenomenon.

\section{Conclusions}

		Three papers have just appeared all claiming that GRB060614 is at z=0.125 and that no associated supernova is detected to severe limits, thus forcing a conclusion that "GRB060614 requires a novel explosive process".  The weak point in these claims is that the reported redshift is {\it not} that of the GRB but merely of the brightest galaxy near the line of sight.  The simplest idea is that the galaxy is just a random foreground source unrelated to the burst and that the burst is at a high redshift so as to easily account for the lack of any visible supernova.  Indeed, a detailed calculation shows that roughly one-to-two such false alarms are {\it expected} to date.  Occam's Razor argues strongly against the "new type of massive stellar death" proposal as being an unnecessary new phenomenon that is already well explained by ordinary chance alignment.
		
		But we do have eight ways of getting a redshift to the GRB itself, and that is to use the eight luminosity indicators to derive luminosities and distances.  These tools are already known to work for 69 GRBs with spectroscopic redshifts (see Figure 1), for which the average error in redshift is 26\%.  This confirmed accuracy is more than adequate to determine whether GRB060614 is at z=0.125 or at high redshift where any underlying supernova would be invisible.  We see that the burst light curve is very spiky with many peaks, the spectral lag is near zero, and the $E_{peak}$ value is high; all pointing to a high luminosity event and a high redshift.  Detailed analysis gives a consistent picture from all eight luminosity relations that GRB060614 is really at $z=1.97_{-0.53}^{+0.84}$ and certainly not at $z=0.125$.  This provides a simple and forced explanation as to why no supernova event was visible.  As such, the claim to use GRB060614 as evidence for a "new type of massive stellar death" is certainly wrong.
		
		$~$
		
		We thank David Palmer for information on GRB060505.

\clearpage

\begin{deluxetable}{lll}
\tabletypesize{\scriptsize}
\tablecaption{Luminosity Relations and Their Derived Luminosities for GRB060614
\label{tbl1}}
\tablewidth{0pt}
\tablehead{
\colhead{Luminosity Relations}   &
\colhead{For $z=0.125$}   &
\colhead{For $z=1.97$ }
}
\startdata
$\log L = 51.25-1.01( \log [\tau_{lag}/(1+z)])$				&
	$\log L = 53.02 \pm 0.60$							&
	$\log L = 53.45 \pm 0.60$  \\
$\log L = 55.50+1.77( \log [V(1+z)])$						&
	$\log L = 51.52 \pm 0.53$							&
	$\log L = 52.26 \pm 0.51$ \\
$\log L = 48.05+1.68( \log [E_{peak}(1+z)])$				&
	$\log L = 52.30 \pm 0.47$							&
	$\log L = 53.00 \pm 0.48$  \\
$\log E_{\gamma ,iso} = 48.65+1.75( \log [E_{peak}(1+z)])$	&
	$\log E_{\gamma ,iso} = 53.08 \pm 0.31$				&
	$\log E_{\gamma ,iso} = 53.82 \pm 0.41$  \\
$\log E_{\gamma} = 46.53+1.63( \log [E_{peak}(1+z)])$		&
	$\log E_{\gamma} = 50.65 \pm 0.31$				&
	$\log E_{\gamma} = 51.34 \pm 0.31$  \\
$\log L = 51.33-1.21( \log [\tau_{RT,min}/(1+z)])$			&
	$\log L = 52.87 \pm 0.49$							&
	$\log L = 53.38 \pm 0.49$  \\
$\log L > 50.32 +2( \log [N_{peak}])$ for $N_{peak} \ge 2$	&	$\log L > 52.23$								&	$\log L > 52.23$  \\
$\log L = 48.50+1.62( \log [E_{peak}(1+z)])-0.49( \log [T_{0.45}(1+z)^{-0.6}])$   &
	$\log L = 52.00 \pm 0.37$							&
	$\log L = 52.80 \pm 0.38$  \\
\enddata
    
\end{deluxetable}

\clearpage

\begin{figure}
\epsscale{.80}
\plotone{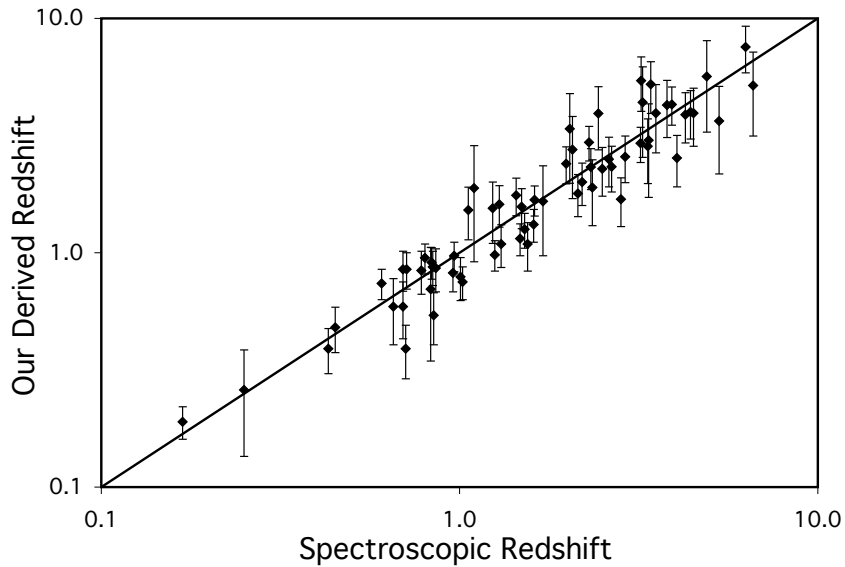}
\caption{
Derived redshifts for 69 GRBs.  We have previously measured the luminosity indicators for 69 GRBs with spectroscopic redshifts, calibrated the luminosity relations, combined the derived distance moduli, and produced a single derived redshift (with one-sigma uncertainties).  This plot shows our derived redshifts versus the spectroscopic redshifts.  If our redshifts were perfect, then they would lie exactly along the diagonal line.  The one-sigma scatter about the diagonal line is 26\% in redshift, with this accuracy being more than adequate to decide whether GRB060614 is at high redshift or at z=0.125.  The reduced chi-square of our derived redshifts compared to the spectroscopic redshifts is close to unity, thus demonstrating that our quoted error bars are realistic.  As such, this diagram is our proof that the GRB luminosity indicators are good tools for determining the redshifts to GRBs.}

\end{figure}

\clearpage

\begin{figure}
\epsscale{.80}
\plotone{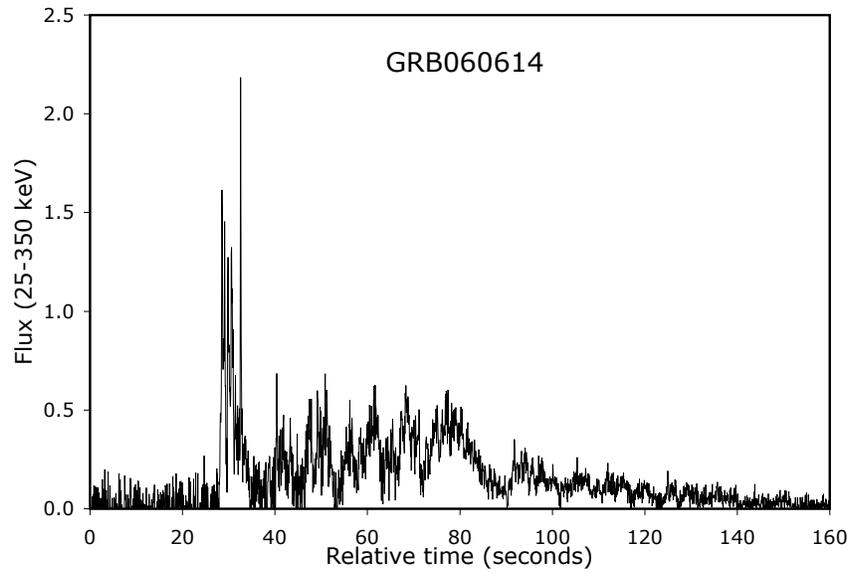}
\caption{
GRB060614 light curve.  The point of this picture is to see the spiky shape, the many peaks, and the very fast rise times; all of which point strongly to a high luminosity event.  With GRB060614 being at high luminosity, it cannot be at z=0.125, and hence the critical link in the argument for a new class of GRBs fails completely.  This light curve is for 25-350 keV with 0.064 second time resolution from the {\it Swift BAT} detector.}

\end{figure}

\end{document}